\documentclass[english,aps,prl,twocolumn,superscriptaddress,showpacs]{revtex4}
\usepackage{graphicx}
\usepackage{babel}
\usepackage{amssymb,amsmath}

\begin{document}

\title{The Implications of Discontinuities for Testing Theories of Turbulence in the Solar Wind}

\author{A.J.~Turner}
\email{a.j.turner@warwick.ac.uk}
\affiliation{Centre for Fusion, Space and Astrophysics; University of Warwick,
Coventry, CV4 7AL, United Kingdom}

\author{S.C.~Chapman}
\affiliation{Centre for Fusion, Space and Astrophysics; University of Warwick,
Coventry, CV4 7AL, United Kingdom}

\author{G.~Gogoberidze}
\affiliation{Physics, Universit\`a della Calabria, Rende, Italy}
\affiliation{Centre for Fusion, Space and Astrophysics; University of Warwick,
Coventry, CV4 7AL, United Kingdom}
\affiliation{Institute of Theoretical Physics, Ilia State University, 3/5 Cholokashvili ave., 0162 Tbilisi, Georgia}

\begin{abstract}
{In-situ} observations of magnetic field fluctuations in the solar wind show a broad continuum in the power spectral density (PSD) with a power-law range of scaling often identified as an inertial range of magnetohydrodynamic turbulence. 
However, both turbulence and discontinuities are present in the solar wind on these inertial range of scales. 
We identify and remove these discontinuities using a method which for the first time does not impose a characteristic scale on the resultant time-series. The PSD of vector field fluctuations obtained from at-a point observations is a tensor that can in principle be anisotropic with scaling exponents that depend on background field and flow direction. This provides a key test of theories of turbulence. We find that the removal of discontinuities from the observed time-series can significantly alter the PSD trace anisotropy. 
It becomes quasi-isotropic, in that the observed exponent does not vary with the background field angle once the discontinuities are removed. This is consistent with the predictions of the Iroshnikov-Kraichnan model of turbulence.
As a consistency check we construct a surrogate time-series from the observations that is composed solely of discontinuities. 
The surrogate provides an estimate of the PSD due solely to discontinuities and this provides the effective noise-floor produced by discontinuities for all scales greater than a few ion-cyclotron scales.

\end{abstract}

\pacs{94.05.Lk, 52.35.Ra, 95.30.Qd, 96.60.Vg}

\maketitle
Turbulence is ubiquitous in astrophysics and may be important in cosmic ray scattering \cite{JC68,C00,BC05}.
\emph{In-situ} satellites provide at-a-point observations of heliospheric plasma parameters over a broad range of temporal scales.
Solar wind observations display fluctuations from plasma kinetic to solar rotation scales \cite{GR99} and exhibit ranges of power-law scaling.
Evidence for Magnetohydrodynamic (MHD) turbulence includes: observed piece-wise power-law scaling in both the components and the trace of the magnetic field power spectral density (PSD) \cite{TM95, H08, H11}, non-Gaussian probability distribution functions (PDF) of the fluctuations that display multi-fractal characteristics \cite{TM95,SV99,H03} and a large estimated effective magnetic Reynolds number \cite{M05,W11}. 
There has as a consequence been considerable interest in using the solar wind to test the predictions of MHD theories of turbulence.

In studies of solar wind turbulence, the models of Iroshnikov-Kraichnan (hereafter IK) \cite{I63, K65} and Goldreich and Sridhar (hereafter GS) \cite{GS95} have merited particular attention (e.g. \cite{C68,Z90,X99,N03,MB03,V07,CH07,H08,A09,P09, S09,L11}).
The isotropic IK model of MHD turbulence predicts $E(k) \sim k^{-3/2}$, whilst the anisotropic model of GS predicts $E(k_\perp) \sim k_\perp^{-5/3}$ and $E(k_\parallel) \sim k_\parallel^{-2}$, where $E(k)$ is the energy spectral density as a function of Fourier wavenumber, k.
The components $k_\perp$ and $k_\parallel$ refer to the orientation of the wavevector w.r.t to a background magnetic field direction, which is often interpreted to be local and scale dependent (e.g. \cite{H08,P09,W10,T12}).
Thus, in principle \emph{in-situ} observations of the solar wind offer a "laboratory" to distinguish between the distinct predictions of IK and GS providing it is possible to identify the background field direction accurately. 
Distinguishing between the predictions of these two theories from data is highly topical (e.g. \cite{C00,H08,P09,W10, GM10,T12}).

In the solar wind there are discontinuities in addition to the turbulence.
Quantifying the effect of discontinuities on statistical measures used in the framework of turbulence is a challenge to data analysis \cite{B01, B04, B08, B10}.
Importantly, discontinuities in a time-series can result in power-law scaling with exponents in the range predicted by turbulence theories \cite{B08}. 
Accurate identification of the background magnetic field direction is critical to testing theories of anisotropic MHD turbulence. However, this can be compromised by the presence of discontinuities.
In this Letter we employ a new method to identify and remove fluctuations affected by discontinuities. 
Discontinuities are usually identified by a change in the magnetic field magnitude and/or direction between two observations that exceeds a given threshold \cite{B08,M11}. 
Uniquely, our method does not assume a single characteristic physical scale for this threshold and hence for discontinuities in the solar wind, and so, does not impose a single scale upon the observed time-series.
This is particularly important when investigating turbulence, which by definition has no characteristic scale.
We find that, once these discontinuities are removed, the remaining signal is consistent with IK but inconsistent with GS, specifically, the trace PSD exponent is isotropic with an exponent close to -3/2 for both $k_\perp$ and $k_\parallel$ sampling directions.

We focus on observations of fast, quiet solar wind using the Ulysses satellite [day 91-146, 1995] whilst at a heliographic latitude of $21^\circ - 58^\circ$ and a radial distance $1.36 - 1.58$ AU.
The interval is recorded at a cadence of $1$ s in an average solar wind flow of $756$ km/s. 
To correct for data gaps and inhomogeneous sampling the data is linearly interpolated to a cadence of $2$ s. 
For periods of missing data longer than $10$ s the resulting region of interpolated data is excluded from subsequent analysis, thus limiting the effect of data gaps and interpolation.

Any study of anisotropic plasma turbulence requires a definition of the background magnetic field direction. 
The background field, $B_0$, is defined by an average over some time-scale $\tau$. 
Previous studies have considered a single time-scale, for example, that of the outer-scale of the turbulence (e.g. \cite{T09}). 
Alternatively, a local background field is defined as a scale dependent estimate over a $\tau$which now varies, where $\tau$ relates the scale of averaging to the scale at which fluctuations are calculated, either using GSF or wavelets (e.g. \cite{H08,P09}).

\begin{figure}
\begin{centering}
\includegraphics[width=.8\columnwidth]{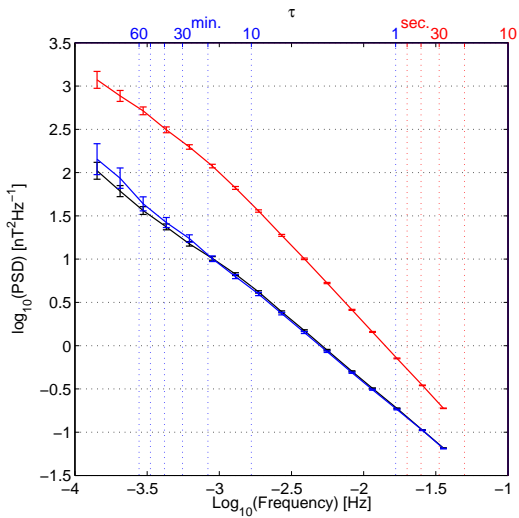}
\par\end{centering}
\caption{ $P^C_{\parallel,R}$ (red), $P^C_{\parallel,F}$ (blue) and $P_{|B|,R}$ (black). C, R, F denote component, raw and discontinuity filtered respectively (See text for details).The spacecraft frequency is on the bottom x-axis with the spacecraft time-scale along the top x-axis with blue and red type for minutes and seconds respectively.}
\label{fig:comp}
\end{figure}

The method we will use is based on an assumption that the magnitude of every vector in the time-series may be written as $B^2 = {\left(B_0 + \epsilon_\parallel \right) }^2 + \epsilon_\perp^2$, where $B_0$ is the background field and $\epsilon_{\perp,\parallel}$ are some small addition to the background field.
Taylor expansion of this expression to lowest order leads to $\delta | \mathbf{B} | = |\mathbf{B}_1| - |\mathbf{B}_2| \sim {B}_{0,1} + \epsilon_{\parallel,1} - {B}_{0,2} - \epsilon_{\parallel,2}$.
Thus, if the background field is unchanged between two observations ${B}_{0,1} = {B}_{0,2}$, then, $\delta | \mathbf{B} | \sim \delta B_\parallel$, where $\delta B_\parallel = \epsilon_{\parallel,1} - \epsilon_{\parallel,2}$. 
Therefore, in the absence of discontinuities in the observed time-series the fluctuation power of the modulus of the magnetic field, $P_{|B|}$, and the power in the parallel component of the fluctuations w.r.t the background field, $P^C_{\parallel}$, will be approximately equal.
This suggests an important criterion for determining whether discontinuities have a significant impact on the statistical properties inferred from the observations.
Note, unlike $P^C_{\parallel}$, $P_{|B|}$ is independent of any averaging scheme used to estimate the background field.
Therefore, the difference between $P^C_{\parallel}$ and $P_{|B|}$ may be regarded as a measure of the quality of the background field estimate.

Here, we use a local background magnetic field estimate, obtained for each scale $\tau$ by averaging the observed time-series with a Gaussian window of width $4 \tau$, where the standard deviation of the Gaussian window determines the spectral characteristics in Fourier space and we may define $f = 1/\tau$ \cite{P09}.
We define the method for detecting discontinuities as follows: a discontinuity is observed at time, $t$, if:
\begin{equation}
| \delta | \mathbf{B} \left( t,f \right)| - \delta B_\parallel \left( t,f \right)| > T_0 f^{-\beta},
\label{eq:thres}
\end{equation}
where $T_0$and $\beta$ are values to be found from the data that determine a scale dependent threshold for the discontinuities.
We have explored a broad range of $T_0$ and $\beta$. 
Here, we show results using $T_0 = 0.022$ and $\beta = 0.61$, which have been selected due to best agreement between $P^C_{\parallel,F}$ and $P_{|B|,R}$, where we refer to the original observations as "raw"(R) and observations filtered by equation (\ref{eq:thres}) as "discontinuity-filtered" (F).

We first show that our procedure removes discontinuities in a consistent fashion. In Fig. \ref{fig:comp} we plot the PSD of  $P^C_{\parallel,R}$ (red), $P^C_{\parallel,F}$ (blue) and $P_{|B|,R}$ (black).
We can see that, $P^C_{\parallel,R} \gg P_{|B|,R}$ (i.e. $\delta B_\parallel \gg \delta|\mathbf{B}|$) and the spectral indices are distinct for time-scales $\tau \lesssim 30$ min.
However, when we consider the discontinuity-filtered case we can see that $P^C_{\parallel,F} \simeq P_{|B|,R}$ within experimental uncertainty.
Agreement between $P^C_{\parallel,F}$ and $P_{|B|,R}$ justifies our choice of $T_0$ and $\beta$.
Thus, we have removed the effect of discontinuities from the observed time-series of the components of the magnetic field in the solar wind according to our definition (\ref{eq:thres}).

When observations are made at a single point in the solar wind it is only possible to infer the reduced one-dimensional energy spectra.
Thus, the observed spectral tensor may be expressed as $\tilde{P}_{ij}(f,\theta) = \int {\rm d}^3 {\bf k} P_{ij}(\mathbf{k}) \delta (2\pi f - \mathbf{k} \cdot \mathbf{V}_{sw})$, where $\tilde{P}_{ij}(f,\theta)$ is the observed spectral tensor, $P_{ij}(\mathbf{k})$ is the three-dimensional spectral tensor, $\mathbf{V}_{SW}$ is the solar wind velocity, $\mathbf{k}$ is the wave vector, $f$ is the frequency in the spacecraft frame and $\theta$ is the angle between the sampling direction and the local background magnetic field \cite{FC76}.
The trace power is then $P^T_\theta(f) = \Sigma_i P_{ii}(f,\theta)$. 
In the case where $\theta$ is subdivided such that $\theta = 80^\circ - 90^\circ \sim \theta_\perp$ and $\theta = 0^\circ - 10^\circ \sim \theta_\parallel$ the corresponding trace power may be regarded as an estimate for the spectra of $k_\perp$ and $k_\parallel$ respectively.
For IK it is expected that $P^T_{\perp,\parallel}(f) \sim f^{-3/2}$ and for GS it is expected that $P^T_\perp(f) \sim f^{-5/3}$ and $P^T_\parallel(f) \sim f^{-2}$.
In the solar wind, $P^T$ has been found to be anisotropic, in that the power and spectral index vary with $\theta$ (e.g. \cite{H08, P09, W10}). 
The scaling is not always consistent with $-5/3$ and $-2$, but generally there is agreement that the spectral indices of $P^T_\perp - P^T_\parallel \sim 1/3$ \cite{W10}. 
These observations are interpreted to support the GS model (See e.g. Refs \cite{H08, P09, W10}).

\begin{figure}
\begin{centering}
\includegraphics[width=\columnwidth]{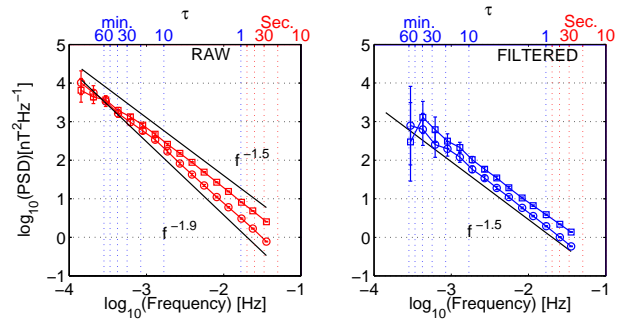}
\par\end{centering}
\caption{ Trace power $P^T_\perp$ and $P^T_\parallel$ shown by squares and circles. Left panel: Raw case. Right panel: Discontinuity-filtered case (See text for details). Examples of power index $-3/2$ and $-1.9$ are shown where appropriate with solid black lines. The spacecraft rest frame frequency is on the bottom x-axis with corresponding time-scale along the top x-axis with blue and red type for minutes and seconds respectively. }
\label{fig:trace}
\end{figure}

We now analyse the raw time-series alongside the discontinuity-filtered time-series to determine how discontinuities affect the measured trace power of the solar wind when sampling parallel, $P^T_\parallel$, and perpendicular, $P^T_\perp$, to the local magnetic field. 
We plot the PSD of the trace power in Fig. \ref{fig:trace}.
In this figure the squares and circles represent the case of $P^T_\perp$ and $P^T_\parallel$.
The left and right panels show the raw and discontinuity-filtered cases respectively.
The solid black lines show fits with PSD exponent $-1.9$ and $-1.5$ where appropriate.

Both $P^T_{\parallel,\perp,F}$ (RH plot) are at lower power than their raw counterparts (LH plot), especially at large scales.
Remarkably, the effect of removing discontinuities is to transform the exponents of the PSD from anisotropic (LH plot) to isotropic (RH plot).
Once discontinuities are removed, the spectral exponents are approximately equal for both $P^T_\parallel$ and $P^T_\perp$, with a value of $-1.5 \pm 0.05$.
This result is dramatically different to that suggested by the raw observations that show that the spectral exponent of $P^T_\parallel$ is greater than $P^T_\perp$, with a difference of $\sim 1/3$ \cite{W10}.
Also, in this interval there is no data above $\tau \sim 1$ hr, implying an upper limit to the time between discontinuities, irrespective of sampling direction w.r.t the background magnetic field.

Our results show that the presence of discontinuities can significantly affect the anisotropy seen in both the power and scaling exponents of the PSD.
Thus, any quantitative comparison between observations and theories of turbulence must address the issue of discontinuities.
Furthermore, using our procedure we obtain a result that supports IK scaling, similar to Ref. \cite{L11}.
Intriguingly, we also find good agreement with a recently propose phenomenological theory of incompressible MHD turbulence \cite{GM10,GM12}.
However, it is not possible to completely confirm this new phenomenology from this study as the parallel and perpendicular dissipation scale must be measured.
This is not possible with the available cadence of the instruments used in this study.

Finally, we will demonstrate that a simple surrogate time-series generated entirely from discontinuities can account for some, but not all, of the power in the \emph{in-situ} observations of fluctuations in the solar wind. 
We will construct a surrogate dataset from the Ulysses fast solar wind observations.’

The surrogate data is a Continuous-Time Markovian jump process, where the angular state space of the observations is preserved in the surrogate time series.
The observations, in $\mathbf{RTN}$ coordinates ( $\mathbf{B}_t = [ B_R(t), B_T(t), B_N(t)]$), are projected onto spherical coordinates ($\mathbf{B}_t = [ |B|_t, \theta_t, \phi_t]$), we may then define the joint probability, $P_{\theta,\phi}$, between angles $\theta$ and $\phi$.
We will construct our surrogate time-series to have the same $P_{\theta,\phi}$ as the original observations.

The waiting time between jumps, $T_W$, is an i.i.d random variable.
The waiting time distribution of the solar wind has been measured by numerous authors \cite{B69, TS79, B01, BC05, V07, B10}.
Reported distributions are exponential, power-law and log-normal.
A close approximation to all of these probability distributions is the Weibull (Stretched-exponential) distribution \cite{CSN09}, such that $T_W(t|\lambda,b) =\frac{b}{\lambda}(\frac{t}{\lambda})^{b-1}\exp(-(\frac{t}{\lambda})^b)$.
Thus, we use the Weibull distribution as our waiting time distribution.
The values for our distribution are chosen by producing the best fit between the PSD of the observations and the surrogate in the time scales, $\tau \sim 1$ hour to $\tau \sim 7$ days.
This is achieved by $ \lambda = 75$ seconds and $b = 0.4$.
With these chosen parameters the mean waiting time is approximately $7$ min.
As our surrogate time series is a jump process the value of our surrogate time series at a specific time, $X_t$, remains constant until a jump occurs.
When a jump occurs a random set from the observations are selected, $\mathbf{B}_j$, where $j$ is a randomly selected index with uniform probability for all data points, which effectively randomises the phase of the discontinuity fluctuations, whilst preserving the observational $P_{\theta,\phi}$.
Our surrogate time series may be written as
\begin{equation}
    X_t=
    \begin{cases}
      [M_0, \theta \left( x_1^{\prime} \right), \phi \left( x_1^{\prime} \right)] , & \text{if}\ t=0 \\
      X_{t-1}, & \text{if}\ t > 0 \text{ and}\ t-s < T_W\\
      [M_0, \theta \left( x_j^{\prime} \right), \phi \left( x_j^{\prime} \right)], & \text{if}\ t > 0 \text{ and}\ t-s = T_W
    \end{cases}
  \end{equation}
where $1/\lambda$ is the jump rate, $t$ is the time of the surrogate data, $s$ is the time of the latest jump in the time series,  $\theta \left( x_j^{\prime} \right)$ and $\phi \left( x_j^{\prime} \right)$ are subsets from the same time index of the observations.
$M_0$ is the constant magnitude of the vector, chosen such that there is agreement between the PSD of the observations and the surrogate date in the range $\tau = 1-7$ days when the surrogate has been projected back to RTN coordinates (these scales are not shown here to improve clarity of the inertial range scales).
The surrogate data is constructed in spherical coordinates, but for the purposes of comparison is plotted in RTN coordinates.

\begin{figure}
\begin{centering}
\includegraphics[width=0.8\columnwidth]{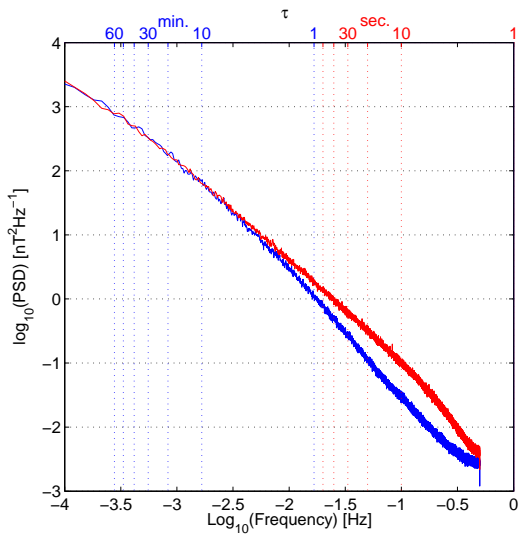}
\par\end{centering}
\caption{ Comparison between the observations (red) and surrogate data of discontinuities (blue) (See text for details). For clarity only the $\mathbf{R}$ component is shown. Spacecraft frequency is along the bottom, whilst spacecraft time-scale is along the top in minutes (blue) and second (red).}
\label{fig:surr}
\end{figure}

We plot the PSD of the surrogate time-series (blue) alongside the observations (red) in Fig. \ref{fig:surr}.
We can see that the observations are closely reproduced by the surrogate at scales $\tau \geq 5$ minutes.
However, where there is no agreement within uncertainty there must also be another process with different autocorrelation properties in the observations distinct from that of the time-series of discontinuities, e.g. turbulence.
This can be seen in Fig. \ref{fig:surr}.
Importantly, we do not claim that only discontinuities are present on time scales $\tau \geq 5$ minutes, only that discontinuities are increasingly the dominant contribution in the observations on these larger scales.

We have presented evidence for more than one process in the solar wind inertial range. 
Discontinuities show increasing PSD dominance as the scales increase but can not account for all the smaller scale fluctuations of the inertial range.
Thus, there must be at least two physical processes present in the inertial range, i.e. discontinuities and turbulence.
The challenge is to separate the PSD contribution from turbulence and discontinuities.
We achieve a separation by using a threshold that does not introduce a characteristic scale to the time-series under investigation.
We measure the difference between two estimates of the parallel component ($\delta B_\parallel$ and $\delta |B|$) to exclude data where the local field is poorly defined/estimated.
We show that our method removes discontinuities, as $P^C_{\parallel,F} \simeq P_{|B|,R}$.
The trace power is shown to to scale in a manner that is independent of $\theta_{BV}$, this is consistent with the IK and more recent quasi-isotropic cascade phenomenology \cite{GM10,GM12}.
Our result does not support the GS model in the solar wind, but suggests previous agreement may be due to a poorly defined local field due to discontinuities in the time-series.
We purpose a new model for discontinuities in the solar wind that demonstrates the contribution to all scales in the coordinate system of the observations.
We find that discontinuities must be considered when using PSD exponents to test theories of turbulence in the solar wind.

\begin{acknowledgments}
The authors acknowledge the Ulysses instrument teams for providing magnetometer data. 
This work was supported by the UK STFC.
\end{acknowledgments}

\end{document}